\documentclass[12pt]{article}
\usepackage{amssymb,amsthm}
\usepackage[cmex10]{amsmath}
\usepackage[longnamesfirst]{natbib}
\usepackage[dvips]{graphicx}
\usepackage{enumerate}
\usepackage{fullpage}
\usepackage{pdfsync}
\usepackage{colordvi}
\usepackage[nodisplayskipstretch]{setspace}
\usepackage{microtype}

\usepackage[T1]{fontenc}

\usepackage{lmodern}

\DeclareMathOperator{\sign}{sign}
\newcommand{\E}{\mathbb{E}}

\newcommand{\IR}{\mathbb{R}}

\newcommand{\1}[1]{\mathbb{I}\{#1\}}
\newcommand{\Var}{\mathbb{V}}

\newcommand{\Cov}{\textrm{Cov}}

\newtheorem{proposition}{Proposition}
\newtheorem{corollary}{Corollary}

\numberwithin{equation}{section}

\usepackage{verbatim,color}
\definecolor{thistle}{rgb}{0.8,0.05,1}

\def\boxit#1{\vbox{\hrule\hbox{\vrule\kern6pt
          \vbox{\kern6pt#1\kern6pt}\kern6pt\vrule}\hrule}}

\DeclareMathOperator*{\argmin}{argmin}

\title{Estimating Derivatives of Function-Valued Parameters in a Class of Moment Condition Models}

\author{ Christoph Rothe and Dominik Wied\mythanks{First Version: October 3, 2016. This Version: \today. Christoph Rothe, Columbia University, Department of Economics, email: {\ttfamily cr2690\@columbia.edu}. Dominik Wied, Institute for Econometrics and Statistics, University of Cologne, email: {\ttfamily dwied@uni-koeln.de}. Research is supported by Deutsche Forschungsgemeinschaft (SFB 823, project A1). }}

\date{}

\usepackage{titlesec}
\titleformat{\section}[block]{\centering\normalfont\sffamily}{\thesection.}{0.5em}{\lsstyle\uppercase}
\titleformat{\subsection}[block]{\normalfont\sffamily}{\thesubsection.}{0.4em plus .1em minus .2em}{}
\titleformat{\subsubsection}[runin]{\normalfont\sffamily}{\thesubsubsection.}{0.4em plus .1em minus .2em}{}[.]
\titlespacing*\section{0pt}{18pt plus 4pt minus 2pt}{4pt plus 1pt minus 1pt}
\titlespacing*\subsection{0pt}{16pt plus 3pt minus 2pt}{4pt plus 1pt minus 1pt}
\titlespacing*\subsubsection{0pt}{12pt plus 2pt minus 1pt}{4pt plus 1pt minus 1pt}

\makeatletter
\def\mythanks#1{%
    \protected@xdef \@thanks {\@thanks \protect \footnotetext [\the \c@footnote ]{#1}}%
}
\makeatother

\begin{document}

\newtheorem{theorem}{Theorem}
\newtheorem{definition}{Definition}
\newtheorem{lemma}{Lemma}
\newtheorem{assumption}{Assumption}
\theoremstyle{definition}
\newtheorem{example}{Example}
\newtheorem{remark}{Remark}

\bibliographystyle{econometrica}

{
\maketitle

\begin{abstract}  We develop a general approach to estimating the derivative of a function-valued
parameter $\theta_o(u)$ that is identified for every value of $u$ as the solution to a moment condition.
This setup in particular covers many interesting models for conditional distributions, such as quantile 
regression or distribution regression. Exploiting that $\theta_o(u)$ solves a moment condition, we obtain
an explicit expression for its derivative from the Implicit Function Theorem, and estimate the components 
of this expression by suitable sample analogues,
which requires the use of (local linear) smoothing. Our estimator can then be used for a variety of 
purposes, including the estimation of conditional density functions, quantile partial effects, and
structural auction models in economics.
\end{abstract}


\noindent \textbf{Keywords:} Quantile Regression, Distribution Regression, Local Linear Smoothing, Conditional Density Estimation, Quantile
Partial Effects
}
\newpage

\doublespacing

\section{Introduction}

Estimating the conditional distribution of a dependent variable $Y$ given covariates $X$ in
$\IR^p$ is an important problem in many areas of applied statistics. In economics, for example,
studies of changes in income inequality often involve estimation of the conditional distribution 
of workers' wages given their observable characteristics \citep[e.g.][]{machado2005cdc,autor2008trends}.
Such applications require models that 
on the one hand are flexible enough to capture the potentially highly heterogeneous impact of 
covariates on the dependent variable at different points in the distribution, but  on the other 
hand can also be estimated using computationally and theoretically attractive methods. A model
that is particularly popular in such contexts is the  linear quantile regression (QR) model 
\citep{koenker1978regression}, which specifies the conditional quantile
function 
$Q_{Y|X}(u,x)$ of $Y$ given $X$ as
\begin{align*}
	Q_{Y|X}(u,x) = x'\theta_o(u).
\end{align*}
Another model that has received much attention recently is the linear distribution regression (DR) model \citep{foresiperacchi:1995},
which specifies the conditional c.d.f.\ $F_{Y|X}(u,x)$ of $Y$ given $X$ as
\begin{align*}
	F_{Y|X}(u,x) = \Lambda(x'\theta_o(u)),
\end{align*}
where $\Lambda(\cdot)$ is a known link function that is often taken to be the Logit function. A common feature
of these two models, and other models for conditional distributions, is that the respective specification
depends on a function-valued parameter $\theta_o(u)$ for which at every appropriate value of $u$ there exist 
an asymptotically normal estimator that converges at the usual parametric rate.

In this paper, we consider the problem of estimating the derivative $\theta_o^u(u) = \partial_u\theta_o(u)$ of
the function-valued parameter in such models.  This derivative plays an important role in estimating many interesting
functionals of a conditional distribution. One application is the
estimation of conditional density functions. In the QR model, for instance, the conditional density $f_{Y|X}(y,x)$
of $Y$ given $X$ is
\begin{align*}
f_{Y|X}(y,x) = \frac{1}{x'\theta_o^u(F_{Y|X}(y,x))}, 
\end{align*}
with $F_{Y|X}(y,x) =\int_0^1 \1{x'\theta_o(u)\leq y}du$ the conditional c.d.f.\ 
of $Y$ given $X$ implied by the QR model. Similarly, in the DR model,
the conditional density of $Y$ given $X$ is 
\begin{align*}
	f_{Y|X}(u,x) =\lambda(x'\theta_o(u))x'\theta_o^u(u),
\end{align*}
with $\lambda(u) = \partial_u\Lambda(u)$ the derivative of the link function. Other applications
we consider in this paper include estimating the distribution of bidders' private valuations of 
auctioned objects based on a QR specification of the distribution of observed bids, and estimating
Quantile Partial Effects (QPEs) in a DR model.

Motivated by these applications, we develop a general approach to estimating the derivative of a function-valued parameter
$\theta_o(u)$ in an abstract class of models in which this parameter is identified for every value of $u$ 
in some index set as the solution to a moment condition or estimating equation. This setup covers both QR
and DR models.  We exploit that $\theta_o(u)$ solves a moment condition to obtain an explicit expression for 
$\theta_o^u(u)$ from the Implicit Function Theorem, and estimate the components of this expression by suitable 
sample analogues. The details of the last step depend on the exact properties of the moment condition, and in both
QR and DR models some form of smoothing is needed. We use local linear smoothing in this case, which leads to a 
computationally simple 
estimator with attractive theoretical properties. For both QR and DR, we show that our estimator of 
$\theta_o^u(u)$ is asymptotically normal and has bias and variance whose order of magnitude is analogous
to that of a one-dimensional nonparametric kernel regression. These properties then carry over to the
above-mentioned applications like density estimation via the Continuous Mapping Theorem.

Our paper is connected to a well-established literature on quantile regression, surveyed
for example in \citet{koenker2005quantile}. It also contributes to an emerging literature
on distribution regression, which was originally proposed by \citet{foresiperacchi:1995} 
and further studied by \citet{chernozhukov2013inference}. See also \citet{rothe2012partial,rothe2015decomp}
for examples of applications of distribution regression in economics, \citet{rothewied:2013}
for specification testing, and \citet{leorato2015comparing} for a comparison with quantile regression.
\citet{chernozhukov2013inference} obtain general results regarding estimation of function-valued 
regular parameters. Our paper seems to be the first to address
the estimation of derivatives of such parameters in general settings, but the specific problem
of estimating the derivative $Q_{Y|X}^u(u,x)=\partial_u Q_{Y|X}(u,x)$ of a conditional quantile 
function with respect to the quantile level has been studied before. In particular, 
\citet{parzen1979nonparametric}, \citet{xiang1995estimation}, and \citet{guerre2012uniform} 
propose methods based on smoothing an estimate of the function $u\mapsto Q_{Y|X}(u,x)$, 
whereas \citet{gimenes2013augmented} propose an estimator based on an augmented quantile 
regression problem with a locally smoothed criterion function. Both approaches differ conceptually
from the one we propose in this paper.

The remainder of this paper is structured as follows. In Section 2, we describe a general approach
to estimating the derivative of the function-valued parameters in a class of models that give rise
to a moment condition or estimating equation of a particular form. In Sections 3 and 4, we apply this
approach to QR and DR models, respectively, and study some applications. Section 5 reports the results
of a simulation study, and Section 6 concludes. All proofs are contained in the appendix.
Throughout the paper, we use  repeated superscripts to denote the partial derivatives 
of functions up to various orders. That is, with $g(y,x)$ a generic function, we write $g^y(y,x)=\partial_y g(y,x)$,
$g^{yy}(y,x)=\partial_y^2 g(y,x)$, $g^{yyy}(y,x)=\partial_y^3 g(y,x)$, etc., for the first, second, third, etc., 
partial derivative with respect to $y$.

\section{General setup}

While we are mostly interested in estimating the derivative of the function-valued parameters
in QR and DR models, we find it useful to motivate our approach in a more general setting that 
also covers other interesting cases. This section describes the approach, obtains some results
on bias properties, and discusses its merits relative to alternative methods.

\subsection{Framework}

We consider a model in which there is a function-valued parameter $u\mapsto\theta_o(u)$, 
with $u\in\mathcal{U}=[u_*,u^*]\subset\IR$ and $\theta_o(u)\in\Theta= \times_{j=1}^p[\theta_{j*},\theta^{j*}]\subset \IR^p$, that is identified 
for every $u\in\mathcal{U}$ through a moment condition. That is, we assume that there exists a function 
$M(\theta,u) = \E(m(Z,\theta,u))$, with $m$ a known function taking values in $\IR^p$ and $Z$ 
an observable random vector, such that 
\begin{align}
	M(\theta,u) =0 \textrm{ if and only if } \theta =\theta_o(u)\label{eq_ident}
\end{align}
for every $u\in\mathcal{U}$. The moment condition $M(\theta,u)$ is assumed to be smooth with respect to
both $\theta$ and $u$, but the underlying function $m(Z,\theta,u)$ can potentially be be non-differentiable. 
We also assume that the data consist of an i.i.d.\ sample $\{Z_i\}_{i=1}^n$ from 
the distribution of $Z$, and  that there is an estimator $\widehat\theta(u)$ of $\theta_o(u)$
satisfying
\begin{align}
	\left\|\widehat{M}(\widehat\theta(u),u)\right\|^2 = \inf_{\theta\in\Theta}\left\|\widehat{M}(\theta,u)\right\|^2 + o_P(n^{-1/2}),
\end{align}
uniformly over $u\in\mathcal{U}$, where $\widehat{M}(\theta,u)=n^{-1}\sum_{i=1}^n m(Z_i,\theta,u)$ is the sample 
version of the moment condition. Under regularity conditions \citep[e.g.][]{chernozhukov2013inference}, the random function 
$u\mapsto \sqrt{n}(\widehat\theta(u) - \theta_o(u))$ then converges to a mean zero Gaussian process with almost 
surely continuous paths, and for fixed $u\in\mathcal{U}$ it holds that
\begin{align}
	\sqrt{n}(\widehat\theta(u) - \theta_o(u))\stackrel{d}{\rightarrow} \mathcal{N}\left(0, M^{\theta}(\theta_o(u),u)^{-1}S(\theta_o(u),u)M^{\theta}(\theta_o(u),u)^{-1}\right),\label{rootn}
\end{align}
where $S(\theta_o(u),u) = \E(m(Z,\theta_o(u),u)m(Z,\theta_o(u),u)^\top)$. Many flexible models for conditional distributions, including QR and DR, fit into this framework.

\subsection{Parameter of interest}

We are interested in estimating the derivative $\theta_o^{u}(u) = \partial_u\theta_o(u)$ of the 
function-valued parameter $\theta_o(u)$, which plays an important role in several applications. Such 
an estimator cannot be obtained by simply taking the (analytical or numerical) derivative of the function
$u\mapsto\widehat\theta(u)$, as this random function is generally not differentiable. The next proposition,
which follows directly from the Implicit Function Theorem, gives primitive conditions for the derivative 
$\theta_o^{u}(u)$ to exist in the first place, and derives an explicit expression.

\begin{proposition}\label{prop1}Suppose that the function $(\theta,u)\mapsto M(\theta,u)$ is $k$ times continuously differentiable
over $\Theta\times\mathcal{U}$, and that the matrix 
$M^{\theta}(\theta,u)$ is invertible. Then the function $u\mapsto \theta_o(u)$ is $k$ times continuously differentiable
over $\mathcal{U}$, and 
\begin{align}
	\theta_o^{u}(u) = -M^{\theta}(\theta_o(u),u)^{-1}M^{u}(\theta_o(u),u). \label{derivrep}
\end{align}
is its first derivative.
\end{proposition}

Heuristically, the formula for $\theta_o^{u}(u)$ can be obtained by taking the total derivative 
of $u\mapsto M(\theta_o(u),u)$, and noting that because of the identification condition~\eqref{eq_ident} 
this derivative is equal to zero for every value of $u$. Analogous reasoning also leads to 
a formula for second- and higher-order derivatives of $\theta_o(u)$, but we do not pursue this
any further in this paper.

\subsection{Estimation approach}
Proposition~\ref{prop1} provides a motivation for constructing an estimator $\widehat{\theta^u}(u)$ of $\theta_o^{u}(u)$
as a sample analogue of the representation~\eqref{derivrep}. That is, with $\widehat{M^{\theta}}(\theta,u)$ and 
$\widehat{M^{u}}(\theta,u)$ suitable sample analogues of ${M^{\theta}}(\theta,u)$ and ${M^{u}}(\theta,u)$,
respectively, we put
\begin{align*}
	\widehat{\theta^{u}}(u) = -\widehat{M^{\theta}}(\widehat\theta(u),u)^{-1} \widehat{M^{u}}(\widehat\theta(u),u).
\end{align*}
This approach is attractive relative to taking the (numerical) derivative of a smoothed version of 
$\widehat\theta(u)$, for example, as it requires computing the estimate $\widehat\theta(u)$ for a 
single value of $u$ only (we discuss the relative merits of the approach more extensively in Section 2.5 below). Moreover, 
obtaining sample analogues of $M^{\theta}(\theta,u)$ and 
$M^{u}(\theta,u)$ is typically not very difficult, although the details depend on the particular 
application. If the function $m(Z,\theta,u)$ is differentiable with respect to $\theta$, we can simply put
\begin{align*}
	\widehat{M^{\theta}}(\theta,u)  = \frac{1}{n}\sum_{i=1}^n m^\theta (Z_i,\theta,u);
\end{align*}
and if $m(Z,\theta,u)$ is differentiable with respect to $u$ we define
\begin{align*}
 \widehat{M^{u}}(\theta,u) = \frac{1}{n} \sum_{i=1}^n m^u (Z_i,\theta,u). 
\end{align*}
Being sample means of simple transformations of i.i.d.\ data, these two quantities
are both easy to compute and straightforward to analyze.

In the QR and DR models that we consider below, however, the function $m(Z,\theta,u)$ is 
only differentiable with respect to one of the two arguments $\theta$ and $u$, and non-differentiable
with respect to the other. Such non-smoothness occurs even though the moment condition $M(\theta,u)$ 
is smooth with respect to both arguments in QR and DR models. This property motivates considering a 
different estimation approach for the case where $m(Z,\theta,u)$ is not smooth. Specifically, if differentiability of 
$m(Z,\theta,u)$ with respect to $u$ fails, we propose to estimate the $j$th component of $M^{\theta}(\theta,u)$ 
by the local linear approximation
\begin{align}
	\widehat{M_j^{u}}(\theta,u)  = \argmin_{\beta\in\IR} \int_{u_*}^{u^*} \left( \widehat M_j(\theta,v) - \widehat M_j(\theta,u) -  \beta(v-u)\right)^2 K_h(v-u) dv, \label{MUdef}
\end{align}
where $K$ is a bounded and symmetric density function with mean zero and compact support, say $[-1,1]$, $h$ is a ``small'' bandwidth 
chosen by the analyst,  $K_h(s)$ is a shorthand notation for $K(s/h)/h$, and $\widehat M_j(\theta,u)$ denotes the $j$th component of the vector
$\widehat M(\theta,u)$. The least squares problem in~\eqref{MUdef} has a unique solution as long as the function $v\mapsto \widehat M_j(\theta,v)$ 
is non-constant over $(u-h,u+h)$. Note that computing this solution does not require the use of numerical optimization methods. Indeed, 
simple algebra shows that
\begin{align*}
	\widehat{M_j^{u}}(\theta,u) &=  \frac{1}{nh\kappa_{2,h}(u) }\sum_{i=1}^n\left( \int_{(u_* -u)/h}^{(u^* -u)/h}  m_j(Z_i,\theta,u+vh)v K(v) dv -  m_j(Z_i,\theta,u) \kappa_{1,h}(u)\right),
\end{align*}
where for any integer $s$ and $u\in\mathcal{U}$ the constant $\kappa_{s,h}(u)$ is defined as $$\kappa_{s,h}(u) = \int_{(u_* -u)/h}^{(u^* -u)/h} v^sK(v)dv.$$
Similarly, if differentiability of $m(Z,\theta,u)$ with respect to $\theta$ fails, we propose to estimate $M^{\theta}(\theta,u)$
by the $(p\times p)$ matrix $\widehat{M^{\theta}}(\theta,u)$ whose $(j,k)$-entry is given by the local linear approximation
\begin{align}
	\widehat{M_{jk}^{\theta}}(\theta,u)  = \argmin_{\beta\in\IR} \int_{\theta_{j*}}^{\theta^{j*}} \left( \widehat M_k(\theta_{-j}(t),u) - \widehat M_k(\theta,u) -  \beta(t-\theta_j)\right)^2 K_h(t-\theta_j) dt.
	\label{MTHETAdef}
\end{align}
 Here $\theta_{-j}(t)=(\theta_1,\ldots, \theta_{j-1},t,\theta_{j+1},\ldots,\theta_p)'$
is a shorthand notation for the $p$-dimensional vector whose $j$th component is equal to $t$, and whose remaining components are equal to the corresponding 
component of $\theta$. As before, the term $\widehat{M_{jk}^{\theta}}(\theta,u)$ can be expressed more explicitly as
\begin{align*}
	&\widehat{M_{jk}^{\theta}}(\theta,u) \\
	&\quad=  \frac{1}{nh\kappa_{2,h}(\theta_j) }\sum_{i=1}^n\left( \int_{(\theta_{j*} -\theta_{j})/h}^{(\theta^{j*} -\theta_{j})/h}  m_k(Z_i,\theta_{-j}(\theta_j +th),u) t K(t) dt
	-  m_k(Z_i,\theta,u) \kappa_{1,h}(\theta_j)\right),
\end{align*}
where for all integers $s,t$ and $\theta\in\Theta$ the constant $\kappa_s(\theta_t)$ is defined as 
$$\kappa_{s,h}(\theta_t) = \int_{(\theta_{t*} -\theta_{t})/h}^{(\theta^{t*} -\theta_{t})/h} v^sK(v)dv.$$
Note that we distinguish the kernel functionals $\kappa_{s,h}(u)$ and $\kappa_{s,h}(\theta_k)$ through the
name of their argument only, which is a slight abuse of notation. Also note that for many models estimation
of $M^{\theta}(\theta,u)$ might already be implemented in many software packages, since such estimates
are needed to construct a plug-in estimator of the asymptotic variance of $\widehat\theta(u)$; see~\eqref{rootn}.

\subsection{Some results on bias under general conditions}

At the current level of generality, it is difficult to conduct a full asymptotic analysis 
of the estimator $\widehat{\theta^u}(u)$ in the ``non-smooth'' case, where either $\widehat{M^{u}}(\theta,u)$
or $\widehat{M^{\theta}}(\theta,u)$  are constructed as described in~\eqref{MUdef} or~\eqref{MTHETAdef},
respectively. The following lemma gives a useful intermediate bias result  under the assumption that 
the moment condition $M(\theta,u)$ satisfies suitable differentiability conditions.

\begin{lemma}Suppose that the function $(\theta,u)\mapsto M(\theta,u)$ is three times continuously differentiable
over $\Theta\times\mathcal{U}$, and that the derivatives are uniformly bounded. Then
\begin{align*}
	\E(\widehat{M^{u}}(\theta,u))-M^{u}(\theta,u) = \frac{h}{2}\frac{\kappa_{3,h}(u)}{\kappa_{2,h}(u)}M^{uu}(\theta,u)+ \frac{h^2}{6}\frac{\kappa_{4,h}(u)}{\kappa_{2,h}(u)}M^{uuu}(\theta,u) +o(h^2)
\end{align*}
if $\widehat{M^{u}}(\theta,u)$ is constructed as described in~\eqref{MUdef}; and
\begin{align*}
	\E(\widehat{M_{jk}^{\theta}}(\theta,u))-M^{\theta}_{jk}(\theta,u) = \frac{h}{2}\frac{\kappa_{3,h}(\theta_j)}{\kappa_{2,h}(\theta_j)}M_{jk}^{\theta\theta}(\theta,u)+ \frac{h^2}{6}\frac{\kappa_{4,h}(\theta_j)}{\kappa_{2,h}(\theta_j)}M_{jk}^{\theta\theta\theta}(\theta,u) +o(h^2)
\end{align*}
if the estimator $\widehat{M^{\theta}_{jk}}(\theta,u)$ is constructed as described in~\eqref{MTHETAdef}.
\end{lemma}

The lemma shows that $\widehat{M_{jk}^{\theta}}(\theta,u)$ has a bias of order $O(h)$ for values of $u$ close to the boundary
of the index set $\mathcal{U}$, and, since $\kappa_{3,h}(u)=0$ for $u\in(u_*+h,u^*-h)$, a bias of order $O(h^2)$ for values
of $u$ sufficiently far in the interior of $\mathcal{U}$. These bias properties are thus similar to those of the Nadaraya-Watson estimator
in a nonparametric regression problem. An analogous statement applies to the elements of the estimator $\widehat{M^{\theta}}(\theta,u)$.
As explained below, however, we can take $\Theta=\IR^p$ in both the QR and the DR model, and thus the bias is of the
order $O(h^2)$ there for all values of $\theta$.
Obtaining further results, such as ones about the  variance of $\widehat{M^{u}}(\theta,u)$ and 
$\widehat{M_{jk}^{\theta}}(\theta,u)$, does not seem possible in non-smooth cases without being more specific
about the nature of the non-smoothness. We therefore only derive such results in the context of our to main
applications, QR and DR, below.

\subsection{Discussion of alternative estimation approaches}

Our approach to estimating $\theta_o^u(u)$ based on the representation~\eqref{derivrep} is by no means
the only possible one. An obvious alternative would be to compute the derivative of a smoothed version
of the function $u\mapsto \widehat\theta(u)$. If local linear smoothing is used for this task, the
$j$th component of the corresponding estimator $\widetilde{\theta^u}(u)$ is given by
\begin{align*}
\widetilde{\theta_j^u}(u) = 	\argmin_{\beta\in\IR} \int \left(\widehat\theta_j(v) - \widehat\theta_j(u) - \beta(v-u)\right)^2K_h(v-u)dv, \quad j=1,\ldots,p.
\end{align*}
This approach is similar  to the ones used by \citet{parzen1979nonparametric}, \citet{xiang1995estimation}, 
and \citet{guerre2012uniform} for estimating  derivatives of quantile functions with respect to the quantile level.
Proceeding like this has the disadvantage that it requires computing $\widehat\theta(u)$ for many values of $u$
over a sufficiently fine mesh in order to approximate the integral with sufficient numerical accuracy, even if 
one is only interested in $\theta_o^u(u)$ for one particular value of $u$.
In contrast, our procedure is computationally much less expensive, as we only require an estimate of $\theta_o(u)$
to estimate $\theta_o^u(u)$.

Another alternative approach is due to \citet{gimenes2013augmented}, who proposed an Augmented Quantile Regression
estimator for the derivative of the function-valued parameter in a QR model. Adapted to our general setting, their
approach amounts to estimating the pair $(\theta_o(u),\theta_o^u(u))$ jointly by solving a linearly augmented
and smoothed version of the moment condition:
\begin{align*}
	\left(\widetilde{\theta}(u), \widetilde{\theta^u}(u)\right) =\argmin_{\theta\in\IR^p,\beta\in\IR^p} \left\| \int \widehat{M}(\theta + \beta(v-u),v)K_h(v-u)dv \right\|^2
\end{align*}
The downside of proceeding like this is that it requires solving a higher-dimensional and slightly
non-standard optimization problem, whereas our estimator can be computed using routines that are implemented
in standard software packages. Moreover, augmented regression as described in the last equation has the disadvantage
that it gives rise to an unnecessary bias term when estimating the function $\theta_o(u)$ itself.

\section{Quantile Regression}

In this section, we study our approach in the context of a QR model,
and consider applications to conditional density and density-quantile estimation,
and to recovering bidders' valuations from auction data.

\subsection{Setup and estimators}

In a linear QR model \citep{koenker1978regression, koenker2005quantile}, the conditional quantile 
function $Q_{Y|X}(u,x)$ of a dependent variable $Y$ given a vector of covariates $X\in\IR^p$ is 
specified for a range of quantile levels $u\in\mathcal{U}=(u_*,u^*)\subset (0,1)$ as $Q_{Y|X}(u,x) = x'\theta_o(u)$,
and the parameter vector $\theta_o(u)$ is estimated by
\begin{align*}
	\widehat\theta(u) = \argmin_{\theta\in\IR^p} \sum_{i=1}^n (u-\1{Y_i \leq X_i'\theta})(Y_i-X_i'\theta).
\end{align*}
Under regularity conditions stated formally below, this model  fits into our general setup with
\begin{align*}
	Z &=(Y,X')', \quad \Theta =\IR^p,\\ m(Z,\theta,u) &= (\1{Y\leq X'\theta}-u)X\\ M(\theta,u) &= \E( (F_{Y|X}(X'\theta,X)-u )X).
\end{align*}
In the QR model, the derivatives of $M(\theta,u)$ with respect to $\theta$ and $u$ are therefore given by
\begin{align*}
	M^\theta(\theta,u) = \E( f_{Y|X}(X_i'\theta,X_i)X_iX_i') \quad\textnormal{ and }\quad
	M^u(\theta,u) = -\E(X_i),
\end{align*}
respectively. Since $M^\theta(\theta,u)$ does not depend on $u$, and $M^u(\theta,u)$ does not 
depend on either $\theta$ or $u$, we denote these objects by $M^\theta(\theta)$ and $M^u$,
respectively, for the remainder of this section to simplify the notation.
We then estimate $M^u$ by
\begin{align*}
	\widehat{M^u} = -\frac{1}{n}\sum_{i=1}^n X_i,
\end{align*}
and construct an estimator $\widehat{M_{jk}^\theta}(\theta)$ of the $(j,k)$ element of $M^\theta(\theta)$ as in described in~\eqref{MTHETAdef}.
The last step yields the expression
\begin{align*}
	\widehat{M_{jk}^\theta}(\theta) = \frac{1}{nh\kappa_2 }\sum_{i=1}^n X_{k,i}\left( \int_{-\infty}^\infty  \1{Y_i\leq X_i'\theta + X_{j,i}th}t K(t) dt \right),
\end{align*}
with $\kappa_s = \int_{-1}^1 v^sK(v)dv$. Note that since $\Theta=\IR^p$, the area of integration in the last equation does not
require a boundary adjustment irrespective of the value of $\theta$. With some algebra, we can write this estimator a bit more
efficiently as
\begin{align*}
		\widehat{M_{jk}^\theta}(\theta)=\frac{1}{n\kappa_2}\sum_{i=1}^n X_{k,i} \sign(X_{j,i}) \bar{K}_h\left(\frac{Y_i- X_i'\theta}{|X_{j,i}|}\right),
\end{align*}
where $\bar{K}(s) = \int_{s}^1 t K(t)dt$ is a new ``pseudo-kernel'' function (it is a symmetric, positive function, but generally does not integrate to one),  $\bar{K}_h(s)= \bar{K}(s/h)/h$, and $\sign(x) = \1{x>0} - \1{x<0}$ is the
sign function. This alternative expression is convenient for the derivation of asymptotic properties, and highlights the similarities
with objects commonly studied in the context of kernel-based nonparametric regression.\footnote{Note that while the matrix $M^\theta(\theta)$
is symmetric under the QR model, the estimator $\widehat{M^{\theta}}(\theta)$ is generally not. To improve finite-sample properties, one
can consider the ``symmetrized'' estimator $(\widehat{M^{\theta}}(\theta)+\widehat{M^{\theta}}(\theta)')/2$ instead. }

The final estimator of the derivative $\theta_o^u(u)$ of the function-valued parameter $\theta_o(u)$ in the QR model is then given by
\begin{align*}
	\widehat{\theta^{u}}(u) = -\widehat{M^{\theta}}(\widehat\theta(u))^{-1} \widehat{M^{u}}.
\end{align*}
To derive the asymptotic properties of $\widehat{\theta^{u}}(u)$, we make the following assumption.

\begin{assumption}\label{QR}
(a) The conditional quantile function takes the form $Q_{Y|X}(u,x) = x'\theta_o(u)$ for all $u\in\mathcal{U}$;
(b) the conditional density function $f_{Y|X}(y,x)$ exists, is uniformly continuous over the support of $(Y,X)$,
uniformly bounded, is twice continuously differentiable with respect to its first argument, and its derivatives 
are uniformly bounded over the support of $(Y,X)$;
(c) The minimal eigenvalue of $M^\theta(\theta)$ is bounded away from zero uniformly over $u\in\mathcal{U}$;
(d) $\E(\|X\|^{4+\delta})<\infty$ for some $\delta>0$;
(e) the bandwidth $h$ satisfies $h\to 0$ and $nh\to\infty$ as $n\to\infty$.
\end{assumption}

Assumption~\ref{QR} collects conditions that are mostly standard in the literature on QR models. Under these conditions, 
both $\widehat\theta_o(u)$ and $\widehat{M^u}$ are $\sqrt{n}$-consistent, whereas each element of the matrix
$\widehat{M^\theta}(\theta_o(u))$ converges to its population counterpart at a slower nonparametric rate. This means that
\begin{align*}
	\widehat{\theta^{u}}(u) - \theta^{u}_o(u) &\cong M^{\theta}(\theta_o(u))^{-1} \left(\widehat{M^{\theta}}(\theta_o(u))-M^{\theta}(\theta_o(u))\right) M^{\theta}(\theta_o(u))^{-1} M^{u},
\end{align*}
and that the stochastic properties of $\widehat{M^\theta}(\theta_o(u))$ drive the asymptotic behavior of $\widehat{\theta^{u}}(u)$.
To state this result formally, we introduce some notation. For every $\theta\in\Theta$, let $\mathbf{A}(\theta)$ be a random $p\times p$ matrix whose elements are jointly normal, have mean
zero, and are such that the  covariance between the $(j,k)$ and the $(l,m)$ element is
\begin{align*}
	&\Cov(\mathbf{A}_{jk}(\theta),\mathbf{A}_{lm}(\theta))  \\
	&\quad = \kappa_2^{-2}\E\left(X_{k,i}X_{m,i} \sign(X_{j,i}X_{l,i}) \int \bar{K}(s)\bar{K}(s |X_{j,i}|/|X_{l,i}|)ds f_{Y|X}(X_i'\theta,X_i )\right).
\end{align*}
The distribution of the random matrix $\mathbf{A}(\theta)$ then implicitly defines a positive-definite matrix $V_o(u)$ that is such that
\begin{align*}
	M^{\theta}(\theta_o(u))^{-1} \mathbf{A}(\theta_o(u))M^{\theta}(\theta_o(u))^{-1} M^{u}\sim\mathcal{N}(0,V_o(u)).
\end{align*}
Note that the matrix $V_o(u)$ could be expressed more explicitly, but this would require notation that is
cumbersome and not very insightful. We also define the bias function
\begin{align*}
	B_o(u) &=M^{\theta}(\theta_o(u))^{-1} A(\theta_o(u))M^{\theta}(\theta_o(u))^{-1} M^{u},
\end{align*}
with $A(\theta)$ the fixed $p\times p$ matrix   whose $(j,k)$ element is equal to
\begin{align*}
	A_{jk}(\theta) =\frac{1}{6}\frac{\kappa_4}{\kappa_2}\E(f^{yy}_{Y|X}(X_i'\theta,X_i) X_{k,i}X_{j,i}^3)
\end{align*}
With this notation, we obtain the following result.

\begin{theorem}Suppose that Assumption~\ref{QR} holds.  Then
\begin{align*}
&\sqrt{nh}(\widehat{\theta^{u}}(u)- \theta^{u}_o(u) - h^2 B_o(u) )\stackrel{d}{\rightarrow}\mathcal{N}(0,V_o(u)).
\end{align*}
\end{theorem}

The theorem shows that $\widehat{\theta^{u}}(u)$ has bias of order $O(h^2)$ and variance of order $O((nh)^{-1})$ for every
value of $u\in\mathcal{U}$. These properties are analogous to those of the local linear estimator in a univariate nonparametric
regression problem. Choosing $h\sim n^{-1/5}$ minimizes the order of the asymptotic mean squared error, and choosing $h$ such
that $nh^5\to 0$ as $n\to \infty$ ensures that the bias of $\widehat{\theta^{u}}(u)$ is asymptotically negligible. In the latter
case, we can also conduct inference using a consistent estimator of the asymptotic variance $V_o(u)$. Such an estimator is
difficult to express explicitly, but can be obtained as follows. First, note that a simple consistent estimator of  the  covariance
between the $(j,k)$ and the $(l,m)$ element of  $\mathbf{A}(\theta_o(u))$ is
\begin{align*}
	\frac{1}{n\kappa_2^2}\sum_{i=1}^n\left(X_{k,i}X_{m,i} \sign(X_{j,i}X_{l,i}) \int \bar{K}(s)\bar{K}(s |X_{j,i}|/|X_{l,i}|)ds \widehat{d}_{Y|X}(u,X_i )\right)
\end{align*}
with $\widehat{d}_{Y|X}(y,x)= 1/x'\widehat{\theta^u}(u)$ the  estimator of the density-quantile function $d_{Y|X}(u,x) \equiv f_{Y|X}(Q_{Y|X}(u,x),x)$
studied in the  subsection after the next one. We can then simulate draws
$\widehat{\mathbf{A}}_s$, $s=1,\ldots,S$, from the distribution of a Gaussian random matrix with mean zero and the just-estimated
covariance structure. Finally, we  obtain an estimate $\widehat V(u)$ of $V_o(u)$ as
\begin{align*}
	\widehat V(u) = \frac{1}{S}\sum_{s=1}^s \widehat T_s \widehat T_s' \quad\textnormal{with}\quad \widehat T_s  =\widehat{M^{\theta}}(\widehat\theta(u))^{-1} \widehat{\mathbf{A}}_s \widehat{M^{\theta}}(\widehat\theta(u))^{-1} \widehat{M^{u}}.
\end{align*}
This estimator is consistent as $S\to\infty$, and can thus be expected to perform reasonably well
if the number of simulation draws $S$ is sufficiently large.

\subsection{Application to density estimation}

We can use the structure implied by a linear QR model to estimate the conditional density function
$f_{Y|X}(y,x)$ of $Y$ given $X$. This is an important application because certain distributional features,
such as the location of modes, are easier to detect on a density graph than on the graph of a quantile function.
In a QR model, we have that
\begin{align*}
f_{Y|X}(y,x) = \frac{1}{x'\theta_o^u(F_{Y|X}(y,x))}, \quad\textnormal{with}\quad F_{Y|X}(y,x) = \int_0^1\1{x'\theta_o(u)\leq y}du
\end{align*}
the conditional c.d.f.\ of $Y$ given $X$ implied by the QR model. By exploiting this structure, we can circumvent
the ``curse of dimensionality'' that makes fully nonparametric estimation of conditional densities infeasible in
settings with many covariates. In particular, we propose the density estimator
\begin{align*}
\widehat{f}_{Y|X}(y,x) = \frac{1}{x'\widehat{\theta^u}(\widehat{F}_{Y|X}(y,x))} \quad\textnormal{with}\quad  \widehat{F}_{Y|X}(y,x)= \epsilon + \int_\epsilon^{1-\epsilon}\1{x'\widehat\theta(u)\leq y}du , 
\end{align*}
for some small constant $\epsilon>0$. Trimming by $\epsilon$ prevents unstable estimates of very low or very high
quantiles to exert an unduly effect on the estimate of the conditional c.d.f. \citep[e.g.][]{chernozhukov2013inference}. Since
$\widehat\theta(u)$ is $\sqrt{n}$-consistent so is $\widehat{F}_{Y|X}(y,x)$, and thus the asymptotic behavior of $\widehat{f}_{Y|X}(y,x)$ is 
driven by that of the slower-converging derivative estimator $\widehat\theta_o^u(u)$ at the point $u=F_{Y|X}(y,x)$.

\begin{corollary}Suppose that Assumption~\ref{QR} holds.  Then
\begin{align*}
	\sqrt{nh}\left(\widehat{f}_{Y|X}(y,x) - f_{Y|X}(y,x)+h^2 \frac{x'B_o(F_{Y|X}(y,x))}{f_{Y|X}(y,x)^{2}}\right) \stackrel{d}{\rightarrow} \mathcal{N}\left(0, \frac{x'V_o(F_{Y|X}(y,x))x}{f_{Y|X}(y,x)^{4}}\right).
\end{align*}
\end{corollary}

We remark that the limiting distribution in the previous corollary can be a poor approximation to the actual finite-sample
distribution of the density estimate $\widehat{f}_{Y|X}$ in areas where the conditional quantile function is rather flat,
and thus $x'\theta_o^u(u)$ is close to zero (both the bias and the asymptotic variance explode in this case).

\subsection{Application to density-quantile estimation}

An application that is closely related to density estimation is that of estimating the density-quantile function 
$d_{Y|X}(u,x) = f_{Y|X}(Q_{Y|X}(u,x),x)$ of $Y$ given $X$. \citet{parzen1979nonparametric} highlights the role of
this function for exploratory data analysis, but it also plays are role for estimating  the asymptotic variance of 
the quantile regression estimator $\widehat\theta(u)$, which is given by $$u(1-u) \E(d_{Y|X}(u,X_i)X_iX_i')^{-1}  \E(X_iX_i' )\E(d_{Y|X}(u,X_i)X_iX_i')^{-1};$$
see \citet{koenker2005quantile}. In the QR model, the density-quantile function and its natural estimator are
easily seen to be
\begin{align*}
d_{Y|X}(u,x) = \frac{1}{x'\theta_o^u(u)} \quad\textrm{and}\quad\widehat{d}_{Y|X}(u,x) = \frac{1}{x'\widehat{\theta^u}(u)}, 
\end{align*}
respectively; and the theoretical properties of the estimator are straightforward to establish. 

\begin{corollary}Suppose that Assumption~\ref{QR} holds.  Then
\begin{align*}
	\sqrt{nh}\left(\widehat{d}_{Y|X}(u,x) - d_{Y|X}(u,x)+h^2 \frac{x'B_o(u)}{d_{Y|X}(u,x)^{2}}\right) \stackrel{d}{\rightarrow} \mathcal{N}\left(0, \frac{x'V_o(u)x}{d_{Y|X}(u,x)^{4}}\right).
\end{align*}
\end{corollary}

Substituting $\widehat{d}_{Y|X}$ for $d_{Y|X}$ in the asymptotic variance formula above, and replacing
expectations with appropriate sample averages, then leads to a new version of the \citet{powell1986censored}
estimator for the asymptotic variance of the quantile regression estimator.

\subsection{Application to estimating bidders' valuations in auctions}

Another interesting way to exploit the structure of a QR model occurs in the analysis of auction data in 
economics. In a first-price sealed-bid auction with independent private values \citep[e.g.][]{guerre2000optimal}, 
an object with observable characteristics $X\in\IR^p$ is auctioned among $b>2$ bidders. Each bidder submits
a bid $Y_j$, $j=1,\ldots,b$, without knowing the bids of the others, and the object is sold to the highest bidder
at the price $\max_{j=1,\ldots,b}Y_j$. Each bidder also has a private (unobserved) valuation $V_j$, $j=1,\ldots,b$ 
for the object, and these valuations are modeled as independent draws from an unknown c.d.f.\ $F_{V|X}(\cdot,X)$.
\citet{guerre2009nonparametric} show that if bidders are risk-neutral the quantiles of the distribution of valuations
can be written in terms of the quantiles of the observed bids as
\begin{align*}
	Q_{V|X}(u,x) = Q_{Y|X}(u,x) +\frac{uQ^{u}_{Y|X}(u|x)}{b-1}.
\end{align*} 
See \citet{haile2003nonparametric}, \citet{marmer2012quantile} and \citet{gimenes2013augmented} for related results.
Using a linear QR specification for the conditional quantile function of observed bids given the object's 
characteristics, we find that 
\begin{align*}
	Q_{V|X}(u,x) = x'\theta(u) +\frac{u x'\theta^{u}(u)}{b-1}.
\end{align*}
A natural estimator of $Q_{V|X}(u,x)$ is thus given by
\begin{align*}
	\widehat{Q}_{V|X}(u|x) = x'\widehat\theta(u) +\frac{u x'\widehat{\theta^u}(u)}{b-1}.
\end{align*}
Since $\widehat\theta(u)$ converges faster than  $\widehat\theta_o^u(u)$, the asymptotic properties of $\widehat{Q}_{V|X}(u,x)$
are again driven by that of the derivative estimator. This is shown formally by the next result.

\begin{corollary}Suppose that Assumption~\ref{QR} holds.  Then
\begin{align*}
	\sqrt{nh}\left(\widehat{Q}_{V|X}(u,x)- Q_{V|X}(u,x) - h^2 \frac{u x'B_o(u)}{b-1}\right) \stackrel{d}{\rightarrow} \mathcal{N}\left( 0, \frac{u^2x'V_o(u)x }{(b-1)^2}\right).
\end{align*}
\end{corollary}

\section{Distribution Regression}

In this section, we study our approach in the context of a DR model,
and consider applications to estimating conditional densities and Quantile Partial Effects (QPEs).

\subsection{Setup and estimators}

In a DR model \citep{foresiperacchi:1995}, the conditional c.d.f. $F_{Y|X}(u,x)$ of  
$Y$ given $X\in\IR^p$ is specified for a range of threshold values $u\in\mathcal{U}=(u_*,u^*)\subset\IR$ as 
$F_{Y|X}(u,x) = \Lambda(x'\theta_o(u))$, where  $\Lambda(\cdot)$ is a known link function. For notational simplicity,
we postulate for this paper that the Logit link $\Lambda(u) = 1/(1+\exp(-u))$ is used, but alternative ones such
as Probit are of course possible as well. For every $u\in\mathcal{U}$, the parameter vector $\theta_o(u)$ is estimated by 
\begin{align*}
	\widehat\theta(u) = \argmin_{\theta\in\IR^p} \sum_{i=1}^n \left(\1{Y_i\leq y}\log(\Lambda(X_i'\theta)) + \1{Y_i > y}\log(1-\Lambda(X_i'\theta)) \right),
\end{align*}
which amounts to fitting a Logistic regression for each $u\in\mathcal{U}$ with $\1{Y_i\leq u}$ as the dependent
variable. Under regularity conditions stated below, this model  fits into our general setup with
\begin{align*}
	Z &=(Y,X')', \quad \Theta =\IR^p,\\ m(Z,\theta,u) &= (\1{Y\leq u} - \Lambda(X'\theta))X\\ M(\theta,u) &= \E\left((F_{Y|X}(u,X_i) - \Lambda(X_i'\theta)) X_i\right).
\end{align*}
In the DR model, the derivatives of $M(\theta,u)$ with respect to $\theta$ and $u$ are therefore given by
\begin{align*}
	M^\theta(\theta,u) = -\E(\lambda(X_i'\theta) X_iX_i') \quad\textnormal{ and }\quad
	M^u(\theta,u) = \E(f_{Y|X}(u,X_i)X_i),
\end{align*}
respectively, where $\lambda(u)=\partial_u\Lambda(u)$ is the derivative of the Logit link function. Since 
$M^\theta(\theta,u)$ does not depend on $u$, and $M^u(\theta,u)$ does not depend on $\theta$, we denote these
objects by $M^\theta(\theta)$ and $M^u(u)$, respectively, for the remainder of this section to simplify the 
notation. 
We then estimate  $M^\theta(\theta)$ by
\begin{align*}
	\widehat{M^\theta}(\theta) = -\frac{1}{n}\sum_{i=1}^n \lambda(X_i'\theta)X_iX_i',
\end{align*}
and  construct an estimator of  $M^u(u)$ as described in~\eqref{MUdef}:
\begin{align*}
	\widehat{M^u}(u) = \frac{1}{nh\kappa_{2,h}(u) }\sum_{i=1}^n X_{i}\left( \int_{u_*}^{u^*}  \1{Y_i\leq u+th}t K(t) dt -\1{Y_i\leq u}\kappa_{1,h}(u)\right).
\end{align*}
This estimator can be written a bit more efficiently as
\begin{align*}
	\widehat{M^u}(u) = \frac{1}{nh \kappa_{2,h}(u)}\sum_{i=1}^n X_{i}\left(\bar{K}\left(\frac{Y_i-u}{h}\right) -\1{Y_i\leq u}\kappa_{1,h}(u)\right),
\end{align*}
where $\bar{K}(s) = \int_{s}^1 tK(t)dt$ as in the previous section; and for values of $u$ such that $u_*+h < u <u^*-h$ we obtain the even simpler representation
\begin{align*}
	\widehat{M^u}(u) = \frac{1}{n\kappa_2}\sum_{i=1}^n X_{i}\bar{K}_h(Y_i-u),
\end{align*}
In any case, we estimate ${\theta^{u}}(u)$ by
\begin{align*}
	\widehat{\theta^{u}}(u) = -\widehat{M^{\theta}}(\widehat\theta(u))^{-1} \widehat{M^{u}},
\end{align*}
and study its asymptotic properties under the following assumption.

\begin{assumption}\label{DR} (a) The conditional c.d.f.\ takes the form $F_{Y|X}(u,x) = \Lambda(x'\theta_o(u))$ for all $u\in\mathcal{U}$;
(b) the conditional density function $f_{Y|X}(y,x)$ exists, is uniformly continuous over the support of $(Y,X)$,
uniformly bounded, is twice continuously differentiable with respect to its first argument, and its derivatives 
are uniformly bounded over the support of $(Y,X)$;
(c) The minimal eigenvalue of $M^\theta(\theta)$ is bounded away from zero uniformly over $u\in\mathcal{U}$;
(d) $\E(\|X\|^{2+\delta})<\infty$ for some $\delta>0$;
(e) the bandwidth $h$ satisfies $h\to 0$ and $nh\to\infty$ as $n\to\infty$.
\end{assumption}

Assumption~\ref{DR} collects conditions that are mostly standard in the literature on DR models. Under these conditions, 
the asymptotic properties of $\widehat{\theta^{u}}(u)$ follow from arguments that are analogous to but simpler than the 
ones used in the context of the QR model in the previous section. In particular, Assumption~\ref{DR} guarantees that both $\widehat\theta_o(u)$
and $\widehat{M^\theta}(\theta)$ are $\sqrt{n}$-consistent, whereas each element of the vector $\widehat{M^u}(u)$ converges
to its population counterpart at a slower nonparametric rate. This means that
\begin{align*}
	\widehat{\theta^{u}}(u) - \theta^{u}_o(u) \cong M^{\theta}(\theta_o(u))^{-1} \left(\widehat{M^u}(u)-M^u(u)\right),
\end{align*}
and that the stochastic properties of $\widehat{M^u}(u)$ drive the asymptotic behavior of $\widehat{\theta^{u}}(u)$. 
To formally state the result, we have to introduce notation that allows us to distinguish the behavior of $\widehat{\theta^{u}}(u)$
for $u$ in the interior and close to the boundary of $\mathcal{U}$. We define the positive-definite variance matrix
\begin{align*}
	V_o(u,c) &= \Gamma(c) \cdot M^{\theta}(\theta_o(u))^{-1}  \E(f_{Y|X}(u,X_i)X_iX_i')  M^{\theta}(\theta_o(u))^{-1},
\end{align*}
where $\Gamma(c) = (\int_{-c}^1 \bar{K}(s)^2ds  + c(\kappa_1(c)^2- 2\kappa_1(c))/\kappa_2(c)^2$ and $\kappa_s(c) = \int^1_{-c}t^sK(t)dt$
are constants that depend on the kernel function only. Note that for $c\geq 1$ the kernel constant $\Gamma(c) = \int_{-1}^1 \bar{K}(s)^2ds /\kappa_2^2$
does not depend on $c$, and we thus write $V_o(u) =V_o(u,1)$.
We also define the bias functions
\begin{align*}
	B_{o,int}(u) = \frac{1}{6}  \frac{\kappa_4}{\kappa_2}\E(f^{uu}_{Y|X}(u,X_i)X_i)\quad\textnormal{and}\quad B_{o,bnd}(c,u) = \frac{1}{2} \frac{\kappa_3(c)}{\kappa_2(c)}\E(f^{u}_{Y|X}(u,X_i)X_i).
\end{align*}
We then obtain the following finding.

\begin{theorem}Suppose that Assumption~\ref{DR} holds.  Then it holds for $u\in\textrm{int}(\mathcal{U})$ that
\begin{align*}
	\sqrt{nh}(\widehat{\theta^{u}}(u) - \theta^{u}_o(u) - h^2 B_{o,int}(u)) \stackrel{d}{\rightarrow}\mathcal{N}\left(0, V_o(u)\right);
\end{align*}
for $u=u_*+ch$ with $c\in(0,1)$ it holds that
\begin{align*}
	\sqrt{nh}(\widehat{\theta^{u}}(u) - \theta^{u}_o(u) - h B_{o,bnd}(u_*,c)) \stackrel{d}{\rightarrow}\mathcal{N}\left(0, V_{o}(u_*,c)\right);
\end{align*}
and for $u=u^*-ch$ with $c\in(0,1)$ it holds that
\begin{align*}
	\sqrt{nh}(\widehat{\theta^{u}}(u) - \theta^{u}_o(u) - h B_{o,bnd}(u^*,c)) \stackrel{d}{\rightarrow}\mathcal{N}\left(0, V_{o}(u^*,c)\right).
\end{align*}
\end{theorem}

The theorem shows that $\widehat{\theta^{u}}(u)$ has bias of order $O(h^2)$ for values of $u$ in the interior of $\mathcal{U}$,
and bias of order $O(h)$ for values of $u$ on the boundary. The  variance of order $O((nh)^{-1})$ for every
value of $u\in\mathcal{U}$. These properties are analogous to those of the Nadaraya-Watson estimator in a univariate nonparametric
regression problem. Choosing $h\sim n^{-1/5}$ minimizes the order of the asymptotic mean squared error $u$ in the interior, and choosing $h$ such
that $nh^5\to 0$ as $n\to \infty$ ensures that the bias of $\widehat{\theta^{u}}(u)$ is asymptotically negligible (and analogously
for values of $u$ on the boundary). In the latter case, we can also conduct inference using a consistent estimator of the asymptotic 
variance $V_o(u)$, such as
\begin{align*}
	\widehat{V}(u) = \frac{\int_{-1}^1 \bar{K}(s)^2ds}{\kappa_2^2} \cdot \widehat{M^{\theta}}(\widehat\theta(u))^{-1} \left(\frac{1}{n}\sum_{i=1}^n\widehat{f}_{Y|X}(u,X_i)X_iX_i'\right) \widehat{M^{\theta}}(\widehat\theta(u))^{-1},
\end{align*}
with $\widehat{f}_{Y|X}(u,x) =\lambda(x'\widehat\theta(u))x'\widehat{\theta^u}(u)$ the density estimator studied in the next subsection.

\subsection{Application to density estimation}

Similarly to the way we used the QR model above, we can use the structure implied by a  DR model to 
estimate the conditional density function $f_{Y|X}(u,x)$ of $Y$ given $X$. The density and its natural 
estimator are given by
\begin{align*}
	f_{Y|X}(u,x) =\lambda(x'\theta_o(u))x'\theta_o^u(u) \quad\textrm{and}\quad \widehat{f}_{Y|X}(u,x) =\lambda(x'\widehat\theta(u))x'\widehat{\theta^u}(u),
\end{align*}
respectively.
Since $\widehat\theta(u)$ is $\sqrt{n}$-consistent and $\widehat\theta_o^u(u)$ converges as a slower rate,
the asymptotic properties of $\widehat{f}_{Y|X}(y,x)$ are driven by those of our derivative estimator.

\begin{corollary}
Suppose that Assumption~\ref{DR} holds.  Then
\begin{align*}
	\sqrt{nh}(\widehat{f}_{Y|X}(u,x) - f_{Y|X}(u,x)-h^2\lambda(x'\theta_o(u))x'B_o(u)) \stackrel{d}{\rightarrow} \mathcal{N}(0, \lambda(x'\theta_o(u))^2 x'V_o(u) x).
\end{align*}
\end{corollary}

Through similar arguments, one could also obtain an estimator of the density-quantile function (see the section on QR models above).
Since this function is less useful in a DR context, we omit the details in the interest of brevity.

\subsection{Application to estimating quantile partial effects}

The vector of Quantile Partial Effects (QPEs) of the conditional distribution of $Y$
given $X$ is formally defined as $\pi(\tau,x) \equiv \partial_x Q_{Y|X}(\tau,x)$ for
any quantile level  $\tau\in(0,1)$. QPEs are widely used and easily interpretable
summary measures in many areas of applied statistics. In the QR model, the function-valued
parameter coincides with the QPE. This means that the parametrization is easily interpretable,
but also imposes the restriction that the function $x\mapsto\pi(\tau,x)$ is constant for every $\tau$.
More flexible nonparametric estimation of QPEs has been considered by \citet{chaudhuri1991global},
\citet{lee2008kernel}, or \citet{guerre2012uniform}; but such methods become practically
infeasible with many covariates due to the ``curse of dimensionality''.

Here we study the use of the DR model as an alternative way to estimate QPEs. This is particularly
attractive in economic application involving wage data, for which \citet{rothewied:2013} argue 
DR often provides a better fit than QR models. An application of the Implicit Function Theorem yields 
that under a DR  specification
\begin{align*}
	\pi(\tau,x) = - \frac{\theta_o(Q_{Y|X}(\tau,x))}{x'\theta_o^u(Q_{Y|X}(\tau,x))}, \quad\textnormal{with}\quad Q_{Y|X}(\tau,x) = \inf\{u: \Lambda(x'\theta_o(u))\geq \tau\}
\end{align*}
the conditional quantile function of $Y$ given $X$ implied by the DR model. This representation of the QPE
suggest the estimator 
\begin{align*}
	\widehat\pi(\tau,x) = - \frac{\widehat{\theta}(\widehat{Q}_{Y|X}(\tau,x))}{x'\widehat{\theta^u}(\widehat{Q}_{Y|X}(\tau,x))}, \quad\textnormal{with}\quad \widehat{Q}_{Y|X}(\tau,x) = \inf\{u: \Lambda(x'\widehat{\theta}(u))\geq \tau\}.
\end{align*}
Since $\widehat\theta(u)$ is $\sqrt{n}$-consistent, so is $\widehat{Q}_{Y|X}(\tau,x)$; and since $\widehat\theta_o^u(u)$ converges as a slower rate,
the asymptotic properties of $\widehat\pi(\tau,x)$ are again driven by those of our derivative estimator.

\begin{corollary}
Suppose that Assumption~\ref{DR} holds.  Then
\begin{align*}
	&\sqrt{nh}\left(\widehat\pi(\tau,x) - \pi(\tau,x) -h^2 \frac{\theta_o(Q_{Y|X}(\tau,x)) x'B_o(Q_{Y|X}(\tau,x))}{(x'\theta_o^u(Q_{Y|X}(\tau,x)))^2}\right)\\
	&\qquad\stackrel{d}{\rightarrow} \mathcal{N}\left(0, \theta_o(Q_{Y|X}(\tau,x))\theta_o(Q_{Y|X}(\tau,x))'\cdot \frac{ x'V_o(Q_{Y|X}(\tau,x))x }{(x'\theta_o^u(Q_{Y|X}(\tau,x)))^4} \right).
\end{align*}
\end{corollary}

\section{Simulations}

To illustrate the finite sample properties of our proposed procedures, we report the results
of a small-scale Monte Carlo study. For brevity, we focus on the QR model.\footnote{Results for the DR model are qualitatively similar,
and are available upon request from the authors.} We generate data as $Y = X + (1+X)U$, where $X$ follows a $\chi^2$ distribution with
1 degree of freedom, $U$ follows a standard Logistic distribution, and $X$ and $U$ are stochastically independent. This means that
the conditional quantile function of $Y$ given $X$ is $Q_{Y|X}(u,x) = \Lambda^{-1}(u) + (\Lambda^{-1}(u) +1)x$, which in turn means
that the linear QR model is correctly specified, with 
\begin{align*}
	\theta_o(u) = \left(\Lambda^{-1}(u), \Lambda^{-1}(u)+1 \right)' \quad\textnormal{and}\quad \theta_o^u(u) = \left(\frac{1}{\Lambda(\lambda(u))}, \frac{1}{\Lambda(\lambda(u))} \right)'.
\end{align*}
We consider estimating $\theta_o^u(u)=(\theta_{o,0}^u(u),\theta_{o,1}^u(u))'$ using the  procedure proposed 
in this paper for the quantile level $u=.5$, sample sizes $n\in\{1000, 4000\}$, and various bandwidth values. 
We also use a triangular kernel, and set the number of replications to 10,000. Results on the estimator's
finite sample bias, variance and mean squared error are given in Table~1. To have a point of reference
for these findings, we also consider estimating $\theta_o^u(u)$ using the two alternative approaches discussed in 
Section 2.5: smoothing the estimated quantile regression process and the using Augmented Quantile Regression 
estimator of \citet{gimenes2013augmented}. The corresponding results are reported in Tables~2 and~3, respectively.

\begin{center}
\textbf{ [TABLES 1--3 ABOUT HERE]}
\end{center}

Overall, our approach compares favorably to the two competing procedures. While the minimal MSE for estimating
the ``intercept'' parameter $\theta_{o,0}^u(u)$ is similar across the three estimator, our procedure has a substantially
smaller MSE when estimating the ``slope'' parameter $\theta_{o,1}^u(u)$. Indeed, MSE is reduced by about one third to one
quarter, depending on  the sample size. This shows the potential usefulness of our proposed procedure for applications,
and shows that it advantages go beyond computational simplicity.
 All estimators are sensitive with respect to the choice of the bandwidth parameter,
and the range of values that produces reasonable results is very different for our procedure than it is for the two
competitors. This is because for our procedure smoothing is with respect to $\theta$, whereas for the two competing procedures
smoothing is with respect to $u$.

\section{Empirical illustration}

In this section, we apply our methods to estimate the conditional density of US workers' wages given
various explanatory variables. The data are taken from 1988 wave of the Current Population Survey (CPS),
an extensive survey of US households. The same data set was previously used in \citet{dinardo1996lmi}, 
to which we refer for details of its construction. It contains information on 74,661 males that were 
employed in the relevant period, including the hourly wage, years of education and years of potential labor
market experience. We fit linear QR and DR models for the conditional distribution 
of the \emph{natural logarithm} of wages given education and experience, and then estimate the corresponding
conditional density function as described above. In Figure~6.1, we plot the result for a worker with 12 years of education
and 16 years of experience, the respective median values of the two variables. For comparison,
we also plot the standard Rosenblatt-Parzen kernel estimator of the density of log-wages, computed
from the only 948 observations in our data with exactly 12 years of education and 16 years of experience.
Both the QR and DR based estimates make use the entire sample, and thus avoid the curse of dimensionality.

	\begin{figure}[!t]
	\begin{center}
\includegraphics[width=\textwidth]{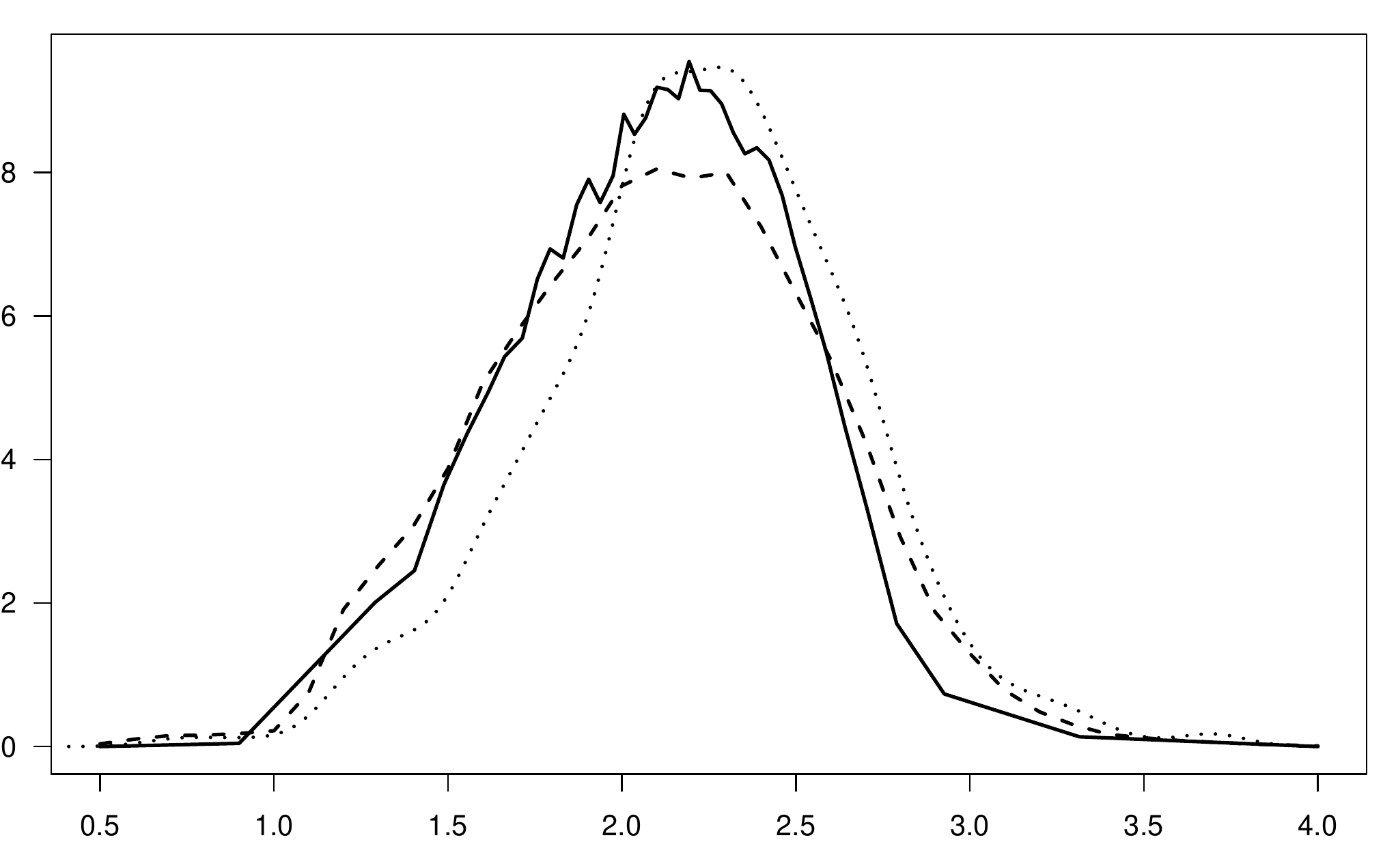}
\caption{\label{fig_s1} Estimated density of the natural logarithm of hourly wages given 12 years of education and 
16 years of experience using a QR specification with $h=.03$ (solid line), DR specification  with $h=.2$ (dashed line) 
and fully nonparametric specification  with $h=.09$ (dotted line).}
\end{center}
\end{figure}

\section{Conclusions}

In this paper, we  propose a new method for estimating the derivative of ``regular'' function-valued parameters
in a class of moment condition models, and provide a detailed analysis of its theoretical properties for the special
cases of Quantile Regression and Distribution Regression models. Possible statistical applications for our method 
include conditional density estimation, estimation of Quantile Partial Effects, and estimation
of auction models in economics. Our simulation results suggests that the method compares favorably to alternative
approaches that have been proposed in the literature.

\appendix

\section{Proofs}

\subsection{Proof of Lemma 1}

We only prove the first statement of the Lemma, as the second one follows from the same type of reasoning.
Using the explicit expression of $\widehat{M_j^{u}}(\theta,u)$ given in the main text, and standard Taylor
expansion arguments commonly used in the kernel smoothing literature, we find that
\begin{align*}
	\E(\widehat{M_j^{u}}(\theta,u)) &=  \frac{1}{h\kappa_{2,h}(u) }\left( \int_{(u_* -u)/h}^{(u^* -u)/h}  M_j(\theta,u+vh)v K(v) dv -  M_j(\theta,u) \kappa_{1,h}(u)\right)\\
	&=\frac{1}{h\kappa_{2,h}(u) }\left( M_j^{u}(\theta,u) h \kappa_{2,h}(u)+ \frac{1}{2} M_j^{uu}(\theta,u) h^2 \kappa_{3,h}(u) \right.\\ &\qquad\left. + \frac{1}{6} M_j^{uuu}(\theta,u) h^3 \kappa_{4,h}(u) + o(h^3)\right)\\
	&=M_j^{u}(\theta,u) + \frac{h}{2} M_j^{uu}(\theta,u) \frac{\kappa_{3,h}(u)}{\kappa_{2,h}(u) }  + \frac{h^2}{6} M_j^{uuu}(\theta,u)\frac{\kappa_{4,h}(u)}{\kappa_{2,h}(u) } +o(h^2),
\end{align*}
as claimed.

\subsection{Proof of Theorem 1}

To simplify the exposition, we prove the Theorem for the special case that all components of $X$ only take
strictly positive values, with probability 1 (the general result follows from the same arguments with an additional
case distinction). We begin by studying the properties of the matrix $\widehat{M^\theta}(\theta)$. Simple algebra shows
that its $(j,k)$ element is
\begin{align*}
	\widehat{M^\theta_{jk}}(\theta) &= \frac{1}{nh\kappa_2 }\sum_{i=1}^n X_{k,i}\left( \int  \1{Y_i\leq X_i'\theta_{-j}(\theta_j+th)}t K(t) dt \right)\\
	&=\frac{1}{nh\kappa_2 }\sum_{i=1}^n X_{k,i}\left( \int  \1{Y_i\leq X_i'\theta + X_{j,i}th}t K(t) dt \right)\\
	&=\frac{1}{nh\kappa_2 }\sum_{i=1}^n X_{k,i}\left( \int_{t\geq (Y_i- X_i'\theta )/(X_{j,i}h)}  t K(t) dt \right)\\
	&=\frac{1}{nh\kappa_2 }\sum_{i=1}^n X_{k,i}\bar{K}\left(\frac{Y_i- X_i'\theta}{X_{j,i}h}\right)
\end{align*}
where $ \bar{K}(s) = \int_{s}^1 t K(t)dt$. Standard kernel calculations involving a change of variables and a Taylor expansion
of the conditional p.d.f.\ $f_{Y|X}$ then yield that
\begin{align*}
	\E\left(\widehat{M^\theta_{jk}}(\theta)\right) = M^\theta_{jk}(\theta) + \frac{h^2}{6}\frac{\kappa_4}{\kappa_2} \E(f^{yy}_{Y|X}(X_i'\theta,X_i) X_{k,i}X_{j,i}^3) + o(h^2);
\end{align*}
so that $\widehat{M^\theta_{jk}}(\theta)$ has bias of order $O(h^2)$. Similarly, we have that
\begin{align*}
	\E\left(X_{k,i}^2 \bar{K}\left(\frac{Y_i- X_i'\theta}{X_{j,i}h}\right)^2\right) &= h\E\left(X_{k,i}^2  \int \bar{K}(s)^2 f_{Y|X}(X_{j,i}sh + X_i'\theta,X_i )ds\right)\\
	&=h \int \bar{K}(s)^2ds \E\left(X_{k,i}^2  f_{Y|X}(X_i'\theta,X_i )\right) + o(h) \quad\textnormal{and}\\
	\E\left(X_{k,i} \bar{K}\left(\frac{Y_i- X_i'\theta}{X_{j,i}h}\right)\right) &= h\E\left(X_{k,i}  \int \bar{K}(s) f_{Y|X}(X_{j,i}sh + X_i'\theta,X_i )ds\right)\\
	&=h \int \bar{K}(s)ds \E(X_{k,i} f_{Y|X}(X_i'\theta,X_i )) + o(h);
\end{align*}
which means that $\widehat{M^\theta_{jk}}(\theta)$ has variance of order $O((nh)^{-1})$:
\begin{align*}
	\Var\left(\widehat{M^\theta}_{jk}(\theta)\right) &= \frac{1}{nh}\frac{\int \bar{K}(s)^2ds}{\kappa_2^2} \E(X_{k,i}^2  f_{Y|X}(X_i'\theta,X_i )) + o((nh)^{-1}).
\end{align*}
Note that the leading term in this variance does not depend on $j$. Next, we calculate the covariance between $\widehat{M^\theta_{jk}}(\theta)$ and 
$\widehat{M^\theta_{lm}}(\theta)$. Since, by the smoothness properties of the conditional density function $f_{Y|X}$, we have that
\begin{align*}
	&\E\left(X_{k,i}X_{l,i} \bar{K}\left(\frac{Y_i- X_i'\theta}{X_{j,i}h}\right) \bar{K}\left(\frac{Y_i- X_i'\theta}{X_{m,i}h}\right)\right) \\
	&\quad=h\E\left(X_{k,i}X_{l,i} \int \bar{K}(s)\bar{K}(s X_{j,i}/X_{m,i})f_{Y|X}(X_{j,i}sh + X_i'\theta,X_i )\right)ds\\
	&\quad=h\E\left(X_{k,i}X_{l,i} \int \bar{K}(s)\bar{K}(s X_{j,i}/X_{m,i})ds f_{Y|X}(X_i'\theta,X_i )\right) +o(h),
\end{align*}
we find that
\begin{align*}
	&\Cov\left(\widehat{M^\theta}_{jk}(\theta),\widehat{M^\theta}_{lm}(\theta)\right)  \\ &\quad= \frac{1}{nh\kappa_2^2}\E\left(X_{k,i}X_{l,i} \int \bar{K}(s)\bar{K}(s X_{j,i}/X_{m,i})ds f_{Y|X}(X_i'\theta,X_i )\right) + o((nh)^{-1})
\end{align*}
It also follows from Lyapunov's central limit theorem and the restrictions on the bandwidth that the joint distribution
of the
$$\sqrt{nh}\left(\widehat{M^\theta}_{jk}(\theta) - M^\theta_{jk}(\theta)\right), \quad (j,k) \in\{1,\ldots,p\}^2$$
is asymptotically (as $n\to\infty$) multivariate normal, with the covariance structure given in the main part of the paper.
From \citet{chernozhukov2013inference}, it follows that $\widehat\theta(u) = \theta_o(u) + O_P(n^{-1/2})$ uniformly over $u\in\mathcal{U}$,
which means that $$\widehat{M^\theta}_{jk}(\widehat\theta(u)) = \widehat{M^\theta}_{jk}(\theta_o(u)) + o((nh)^{-1});$$
and we clearly have that $\widehat{M^{u}}= M^u + O_P(n^{-1/2})$. From a first-order Taylor expansion of the inverse of a matrix, we then
get that
\begin{align*}
	\widehat{\theta^{u}}(u) &= -\widehat{M^{\theta}}(\widehat\theta(u))^{-1} \widehat{M^{u}}\\
	& = -\widehat{M^{\theta}}(\theta_o(u))^{-1} M^{u} + o_p(n^{-1/2})\\
	&= - M^{\theta}(\theta_o(u))^{-1} \left(\widehat{M^{\theta}}(\theta_o(u))-M^{\theta}(\theta_o(u)) \right) M^{\theta}(\theta_o(u))^{-1} M^{u}\\
	&\qquad -M^{\theta}(\theta_o(u))^{-1} M^{u}  + o_P(n^{-1/2}).
\end{align*}
Noting that $\theta^{u}(u) = -M^{\theta}(\theta_o(u))^{-1} M^{u}$, we then obtain the statement of the Theorem. \qed

\subsection{Proof of Theorem 2}

The proof of this theorem follows from arguments that are similar to those used to proof Lemma~1 and Theorem~2,
but substantially simpler. We thus omit the details for brevity.

\bibliography{bibl}

\newpage

\begin{table}[!h]
\centering \caption{Simulation results using estimator proposed in this paper}
\begin{tabular}{cccccccccc}
  \hline\hline
	&&& \multicolumn{2}{c}{Bias}& \multicolumn{2}{c}{Variance}& \multicolumn{2}{c}{MSE}\\
  $u$ & $n$ & $h$ & $\widehat\theta_{o,0}^u(u)$ &  $\widehat\theta_{o,1}^u(u)$ &  $\widehat\theta_{o,0}^u(u)$ &  $\widehat\theta_{o,1}^u(u)$ &  $\widehat\theta_{o,0}^u(u)$ &  $\widehat\theta_{o,1}^u(u)$ \\ 
  \hline
 .5 & 1000&   0.5 & -0.180 & -0.001 & 0.490 & 1.038 & 0.522 & 1.038 \\ 
 && 0.7 & -0.110 & 0.022 & 0.328 & 0.706 & 0.340 & 0.707 \\ 
  &&0.9 & -0.053 & 0.044 & 0.267 & 0.523 & 0.270 & 0.525 \\ 
 && 1.1 & -0.010 & 0.077 & 0.208 & 0.415 & 0.208 & 0.421 \\ 
 && 1.3 & 0.044 & 0.093 & 0.174 & 0.328 & 0.176 & 0.336 \\ 
 && 1.5 & 0.089 & 0.130 & 0.147 & 0.267 & 0.155 & 0.283 \\ 
&&  1.7 & 0.131 & 0.165 & 0.128 & 0.224 & 0.145 & 0.251 \\ 
&&  2.0 & 0.208 & 0.221 & 0.114 & 0.182 & 0.157 & 0.231 \\ 
 && 3.0& 0.490 & 0.485 & 0.080 & 0.104 & 0.320 & 0.339 \\\hline 
.5 & 4000&   0.5 & -0.028 & 0.010 & 0.121 & 0.245 & 0.122 & 0.245 \\ 
 && 0.7 & 0.003 & 0.022 & 0.084 & 0.170 & 0.084 & 0.171 \\ 
 && 0.9 & 0.026 & 0.048 & 0.063 & 0.127 & 0.064 & 0.130 \\ 
 && 1.1 & 0.057 & 0.068 & 0.052 & 0.097 & 0.055 & 0.102 \\ 
 && 1.3 & 0.097 & 0.094 & 0.042 & 0.080 & 0.052 & 0.089 \\ 
 && 1.5 & 0.133 & 0.129 & 0.037 & 0.067 & 0.055 & 0.083 \\ 
 && 1.7 & 0.171 & 0.167 & 0.032 & 0.056 & 0.061 & 0.084 \\ 
 && 2.0 & 0.246 & 0.227 & 0.027 & 0.045 & 0.087 & 0.097 \\ 
 && 3.0 & 0.522 & 0.486 & 0.019 & 0.026 & 0.292 & 0.262 \\    \hline\hline
\end{tabular}
\end{table}

\newpage

\begin{table}[!h]
\centering \caption{Simulation results using smoothed quantile regression coefficients}
\begin{tabular}{cccccccccc}
  \hline\hline
	&&& \multicolumn{2}{c}{Bias}& \multicolumn{2}{c}{Variance}& \multicolumn{2}{c}{MSE}\\
  $u$ & $n$ & $h$ & $\widehat\theta_{o,0}^u(u)$ &  $\widehat\theta_{o,1}^u(u)$ &  $\widehat\theta_{o,0}^u(u)$ &  $\widehat\theta_{o,1}^u(u)$ &  $\widehat\theta_{o,0}^u(u)$ &  $\widehat\theta_{o,1}^u(u)$ \\ 
  \hline
 .5 & 1000 &0.05 & 0.011 & 0.010 & 0.794 & 2.297 & 0.794 & 2.298 \\ 
 && 0.10 & 0.030 & 0.021 & 0.368 & 1.053 & 0.369 & 1.053 \\ 
&&  0.15 & 0.057 & 0.047 & 0.236 & 0.678 & 0.239 & 0.680 \\ 
 && 0.20 & 0.099 & 0.084 & 0.172 & 0.497 & 0.182 & 0.504 \\ 
 && 0.25 & 0.155 & 0.137 & 0.135 & 0.390 & 0.159 & 0.409 \\ 
 && 0.30 & 0.228 & 0.209 & 0.111 & 0.319 & 0.163 & 0.363 \\ 
 && 0.35 & 0.325 & 0.305 & 0.095 & 0.272 & 0.200 & 0.365 \\ 
 && 0.40 & 0.453 & 0.433 & 0.083 & 0.239 & 0.289 & 0.426 \\ 
 && 0.50 & 0.922 & 0.892 & 0.072 & 0.204 & 0.921 & 1.000 \\ \hline
 .5 & 4000 &  0.05 & 0.006 & 0.002 & 0.196 & 0.559 & 0.196 & 0.559 \\ 
&&  0.10 & 0.022 & 0.018 & 0.090 & 0.256 & 0.091 & 0.256 \\ 
&&  0.15 & 0.050 & 0.044 & 0.058 & 0.163 & 0.060 & 0.165 \\ 
&&  0.20 & 0.091 & 0.085 & 0.042 & 0.118 & 0.050 & 0.125 \\ 
&&  0.25 & 0.147 & 0.141 & 0.033 & 0.092 & 0.054 & 0.112 \\ 
&&  0.30 & 0.220 & 0.214 & 0.027 & 0.076 & 0.075 & 0.122 \\ 
&&  0.35 & 0.316 & 0.311 & 0.023 & 0.065 & 0.123 & 0.162 \\ 
&&  0.40 & 0.443 & 0.440 & 0.020 & 0.057 & 0.217 & 0.251 \\ 
&&  0.50 & 0.909 & 0.903 & 0.017 & 0.049 & 0.843 & 0.865 \\     \hline\hline
\end{tabular}
\end{table}

\newpage

\begin{table}[!h]
\centering \caption{Simulation results using augmented quantile regression}
\begin{tabular}{cccccccccc}
  \hline\hline
	&&& \multicolumn{2}{c}{Bias}& \multicolumn{2}{c}{Variance}& \multicolumn{2}{c}{MSE}\\
  $u$ & $n$ & $h$ & $\widehat\theta_{o,0}^u(u)$ &  $\widehat\theta_{o,1}^u(u)$ &  $\widehat\theta_{o,0}^u(u)$ &  $\widehat\theta_{o,1}^u(u)$ &  $\widehat\theta_{o,0}^u(u)$ &  $\widehat\theta_{o,1}^u(u)$ \\ 
  \hline
 .5 & 1000 & 0.05 & 0.035 & -0.321 & 0.776 & 2.182 & 0.778 & 2.285 \\ 
 && 0.10 & 0.043 & -0.146 & 0.372 & 1.047 & 0.373 & 1.069 \\ 
 && 0.15 & 0.067 & -0.064 & 0.241 & 0.685 & 0.245 & 0.689 \\ 
 && 0.20 & 0.104 & 0.003 & 0.178 & 0.509 & 0.188 & 0.509 \\ 
 && 0.25 & 0.155 & 0.071 & 0.142 & 0.406 & 0.166 & 0.411 \\ 
 && 0.30 & 0.220 & 0.148 & 0.119 & 0.339 & 0.167 & 0.361 \\ 
 && 0.35 & 0.302 & 0.238 & 0.103 & 0.295 & 0.194 & 0.351 \\ 
 && 0.40 & 0.401 & 0.345 & 0.092 & 0.263 & 0.253 & 0.382 \\ 
 && 0.50 & 0.672 & 0.624 & 0.081 & 0.231 & 0.533 & 0.620 \\ \hline
 .5 & 4000 &  0.05 & 0.013 & -0.083 & 0.190 & 0.535 & 0.190 & 0.542 \\ 
 && 0.10 & 0.025 & -0.024 & 0.090 & 0.254 & 0.091 & 0.255 \\ 
 && 0.15 & 0.052 & 0.017 & 0.059 & 0.164 & 0.061 & 0.164 \\ 
 && 0.20 & 0.091 & 0.064 & 0.043 & 0.121 & 0.051 & 0.125 \\ 
 && 0.25 & 0.143 & 0.120 & 0.034 & 0.096 & 0.055 & 0.111 \\ 
&&  0.30 & 0.209 & 0.189 & 0.029 & 0.080 & 0.072 & 0.116 \\ 
 && 0.35 & 0.290 & 0.274 & 0.025 & 0.070 & 0.109 & 0.145 \\ 
 && 0.40 & 0.390 & 0.377 & 0.022 & 0.063 & 0.175 & 0.205 \\ 
 && 0.50 & 0.661 & 0.650 & 0.020 & 0.056 & 0.456 & 0.479 \\     \hline\hline
\end{tabular}
\end{table}

\end{document}